\documentclass[referee,sn-nature]{sn-jnl}

\usepackage{graphicx}%
\usepackage{multirow}%
\usepackage{amsmath,amssymb,amsfonts}%
\usepackage{amsthm}%
\usepackage{mathrsfs}%
\usepackage[title]{appendix}%
\usepackage{xcolor}%
\usepackage{textcomp}%
\usepackage{manyfoot}%
\usepackage{booktabs}%
\usepackage{algorithm}%
\usepackage{algorithmicx}%
\usepackage{algpseudocode}%
\usepackage{listings}%
\usepackage{changes}

\usepackage{geometry}
\theoremstyle{thmstyleone}%
\theoremstyle{thmstyletwo}%

\theoremstyle{thmstylethree}%

\begin{document}

\title[Article Title]{Direct observation of Floquet-Bloch states in monolayer graphene}


\author[1]{\fnm{Dongsung} \sur{Choi}}
\equalcont{These authors contributed equally to this work.}
\author[2,3]{\fnm{Masataka} \sur{Mogi}}
\equalcont{These authors contributed equally to this work.}
\author[4,5]{\fnm{Umberto} \sur{De Giovannini}}
\author[2]{\fnm{Doron} \sur{Azoury}}
\author[2,6]{\fnm{Baiqing} \sur{Lv}}
\author[2]{\fnm{Yifan} \sur{Su}}
\author[4]{\fnm{Hannes} \sur{H\"ubener}}
\author[4,7]{\fnm{Angel} \sur{Rubio}}
\author*[2]{\fnm{Nuh} \sur{Gedik}}\email{gedik@mit.edu}

\affil[1]{\orgdiv{Department of Electrical Engineering and Computer Science}, \orgname{Massachusetts Institute of Technology}, \orgaddress{\city{Cambridge}, \state{MA} \postcode{02139}, \country{United States}}}
\affil[2]{\orgdiv{Department of Physics}, \orgname{Massachusetts Institute of Technology}, \orgaddress{\city{Cambridge}, \state{MA} \postcode{02139}, \country{United States}}}
\affil[3]{\orgdiv{Department of Applied Physics}, \orgname{University of Tokyo}, \orgaddress{\street{Bunkyo-ku}, \city{Tokyo}, \country{Japan}}}
\affil[4]{\orgdiv{Center for Free-electron Laser Science}, \orgname{Max Planck Institute for the Structure and Dynamics of Matter}, \orgaddress{\city{Hamburg} \postcode{22761}, \country{Germany}}}
\affil[5]{\orgdiv{Dipartimento di Fisica e Chimica - Emilio Segr\`{e}}, \orgname{Universit\`{a} degli Studi di Palermo}, \orgaddress{\city{Palermo} \postcode{I-90123}, \country{Italy}}}
\affil[6]{\orgdiv{School of Physics and Astronomy}, \orgname{Shanghai Jiao Tong University}, \orgaddress{\city{Shanghai}, \country{People’s Republic of China}}}
\affil[7]{\orgdiv{Center for Computational Quantum Physics (CCQ)}, \orgname{The Flatiron Institute}, \orgaddress{\city{New York}, \state{New York} \postcode{10010}, \country{United States}}}


\abstract{
Floquet engineering is a novel method of manipulating quantum phases of matter via periodic driving \cite{delaTorre2021,Bao2022}. It has successfully been utilized in different platforms ranging from photonic systems \cite{Rechtsman2013} to optical lattice of ultracold atoms \cite{Jotzu2014, Eckardt2017}.
In solids, light can be used as the periodic drive via coherent light-matter interaction. This leads to hybridization of Bloch electrons with photons resulting in replica bands known as Floquet-Bloch states. After the direct observation of Floquet-Bloch states in a topological insulator \cite{Wang2013}, their manifestations have been seen in a number of other experiments \cite{Sie2014Valley-selectiveMonolayerWS2, Sie2017LargeWS2, McIver2020,Aeschlimann2021, Kim2014UltrafastMonolayers,Shan2021,Park2022,Zhou2023}. By engineering the electronic band structure using  Floquet-Bloch states, various exotic phase transitions have been predicted \cite{Oka2009,Lindner2011,Lindner2013,Wang2014,Mentink2015,Ebihara2016,Chan2016,Huebener2017} to occur. To realize these phases, it is necessary to better understand the nature of Floquet-Bloch states in different materials. However, direct energy and momentum resolved observation of these states is still limited to only few material systems \cite{Wang2013, Mahmood2016,Ito2023Build-upTimescales, Zhou2023, Aeschlimann2021}. Here, we report direct observation of Floquet-Bloch states in monolayer epitaxial graphene which was the first proposed material platform \cite{Oka2009} for Floquet engineering. By using time- and angle-resolved photoemission spectroscopy (trARPES) with mid-infrared (mid-IR) pump excitation, we detected replicas of the Dirac cone. Pump polarization dependence of these replica bands unequivocally shows that they originate from the scattering between Floquet-Bloch states and photon-dressed free-electron-like photoemission final states, called Volkov states. Beyond graphene, our method can potentially be used to directly observe Floquet-Bloch states in other systems paving the way for Floquet engineering in a wide range of quantum materials.
}

\keywords{Floquet-Bloch states, monolayer (epitaxial) graphene, time- and angle-resolved photoemission spectroscopy}



\maketitle

\makeatletter 
\renewcommand{\thesection}{\@arabic\c@section.}
\renewcommand{\thefigure}{\@arabic\c@figure}
\renewcommand{\thetable}{\@arabic\c@table}
\renewcommand{\theequation}{\@arabic\c@equation}
\makeatother

\section{Introduction}\label{introduction}
Ever since the prediction of a Chern insulator state in graphene through manipulation of the electronic band structure with circularly polarized light \cite{Oka2009}, various exotic phase transitions via Floquet engineering have been envisioned in solids. Among others, these include Floquet topological insulator \cite{Lindner2011,Lindner2013}, control of the exchange interaction \cite{Mentink2015}, and Floquet-Weyl semimetal \cite{Wang2014,Ebihara2016,Chan2016,Huebener2017}. Experimentally, Floquet-Bloch states were observed for the first time in the three-dimensional topological insulator, Bi$_{2}$Se$_{3}$ \cite{Wang2013} and the light-induced Chern insulator state was also realized in the same system by breaking time-reversal symmetry with a circularly polarized pump beam \cite{Wang2013}. Subsequently, Floquet driving has been explored in various condensed matter systems such as WSe$_{2}$ \cite{Aeschlimann2021}, MnPS$_{3}$ \cite{Shan2021}, black phosphorous \cite{Zhou2023}, and Cr$_{2}$O$_{3}$ \cite{Zhang2024}. 

\par Floquet engineering has also been intensively investigated in graphene. Ultrafast transport techniques were used \cite{McIver2020} to measure light induced anomalous Hall effect, and superconducting tunneling spectroscopy \cite{Park2022} revealed steady-state Floquet-Andreev states. However, despite various theoretical proposals \cite{Sentef2015,Huebener2018,Schuler2020a,Schuler2020b}, direct observation of Floquet-Bloch states in graphene using trARPES has never been achieved. Several reasons make this a very challenging task: First, unlike topological insulators, the Dirac cone in graphene is located at the K point making it necessary to combine high harmonic generation probe and mid-IR pump excitation in trARPES. Second, even after observation of replica bands in trARPES, a reliable procedure for graphene is lacking to disentangle the contribution of Volkov states. Finally, it was also suggested that high scattering rate in the sample \cite{Sato2020,Aeschlimann2021} might mask their observation. 

\par Here, we report the energy-momentum resolved direct observation of Floquet-Bloch states in monolayer epitaxial graphene using trARPES. After exciting the sample with mid-IR (246 meV) pump pulse, we recorded the evolution of the electronic band structure via photo-emission performed by a synchronized extreme UV (26.4 eV) probe pulse (Fig. \ref{fig_1}a). Measured ARPES spectra show replicas of the Dirac cone indicating the presence of photon dressed states. To identify whether these replicas originate from Floquet-Bloch states, we investigated the evolution of spectral features as the pump polarization was rotated. By using a simple analytical theory inspired by ref. \cite{Park2014}, we show that scattering between Floquet-Bloch and Volkov states (Fig. \ref{fig_1}b) can reproduce the experimental results and Volkov states alone can not account for the observed pump polarization dependence, indicating that Floquet-Bloch states are indeed observed. 

\section{Observation of transient replica bands in monolayer graphene}\label{transient_replica_bands}
Figure \ref{fig_1}c displays a static ARPES spectrum of monolayer epitaxial graphene on SiC obtained in our chamber. It features a well-known Dirac cone at $K$ point. The Dirac point is gapped due to the breaking of A and B sublattice symmetry induced by interaction with the buffer layer \cite{Zhou2007}. The arc-shaped feature of the Fermi surface (Fig. \ref{fig_1}d) is due to photoemission matrix element effects. We excite the monolayer graphene using linearly polarized pump pulses with duration of 250 fs. The pump beam was incident nearly normal to the sample surface. In Fig. \ref{fig_1}f, snapshots depict the emergence and decay of replica bands. There is a depletion in the occupied band and an excitation in the unoccupied band in the original ($n$ = 0) Dirac dispersion, where $n$ corresponds to index of replica bands. This results from multi-photon excitation followed by intraband relaxation and/or intraband carrier acceleration. Nevertheless, we did not detect any avoided crossing gap, since the size of the gap at the Dirac point (approximately 0.27 eV) exceeds the pump photon energy (refer to the section S2 in Supplementary Information (SI)) preventing the crossing of the successive replicas. The absence of the avoided crossing gap leaves the origin of replica bands ambiguous at this stage, as the observation of such a gap would have been conclusive evidence for the existence of Floquet-Bloch states \cite{Park2014}. 

\section{Dependence of replica bands on pump polarization}\label{evolution_kx_ky_cuts}
To elucidate the origin of these replica bands, we investigated the evolution of their spectral features in \textit{k}$_{x}$ - \textit{k}$_{y}$ plane at $E - E_{F} = 0.219$ eV as the pump polarization angle ($\theta_{p}$) is rotated (Fig. \ref{fig_2}a). Here $\theta_{p}$ is defined with respect to $\Gamma - K$ direction. To clearly display the replica band with a weak intensity, panels in Fig. \ref{fig_2}a are normalized such that the color scale is set from the minimum to the maximum value within each panel. At $\theta_{p} = 12.5^{\circ}$, the arc-shaped spectral feature is the $n$ = 1 replica band which is similar to the $n$ = 0 band (Fig. 1d). As we rotate the pump polarization counterclockwise (CCW), we found that the arc also rotates. To follow the intensity distribution in detail, we divided the displayed $\theta_{p}$ range into three segments: range I, middle range, range II, as delineated in Fig. \ref{fig_2}a. In the range I, as $\theta_{p}$ increases, the inner arc rotates CCW. Transitioning into the middle range, at $\theta_{p}$ = 90.6$^{\circ}$, spectral weight emerges at the upper left side. The outer arc becomes more pronounced and it represents the excitation in the unoccupied band in the $n$ = 0 Dirac cone. Subsequently, at $\theta_{p}$ = 98.5$^{\circ}$, the original arc disappears, and a new arc in the upper left side rotates CCW in the range II. To provide a clearer visualization of this evolution, we extracted one-dimensional intensity profiles as a function of the angle around the arc ($\theta_{k}$) which is defined in Fig. \ref{fig_1}e. This analysis reaffirms that as $\theta_{p}$ increases, the profile shifts towards the right, corresponding to a CCW rotation in both ranges I and II. Furthermore, it clarifies the evolution of the two arcs within the middle range: two peaks (or arcs) coexist, and with increasing $\theta_{p}$, the primary peak in the range I diminishes while a new peak emerges on the left side, becoming the primary peak in the range II. These observations prompt the question: are Floquet-Bloch states necessary to explain these features, or can they be understood solely based on Volkov states?

\section{Theoretical modeling of the pump polarization dependent replica bands}\label{theoretical_model}
To understand the evolution of replica bands with pump polarization, we employed a model describing the probability density of photoelectron states using the scattering matrix elements derived in \cite{Park2014}. We conducted simulations for three distinct cases: the scattering between Floquet-Bloch and Volkov states (Floquet-Volkov), only Volkov states, and only Floquet states. According to ref. \cite{Park2014}, the photoelectron probability (squared modulus of the scattering matrix element) of the $n$ = 1 replica state is $P_{1} \propto \lvert M \rvert^2 \times \lvert \gamma \rvert^2$, where $M$ denotes the matrix element of the photoemission process from the ground state ($\lvert M \lvert^2 \sim \sin^2(\theta_k/2)$ in monolayer graphene \cite{Hwang2011}). $\lvert \gamma \rvert^2$ characterizes the strength of sidebands, containing scattering processes between the Floquet-Bloch and Volkov states as presented by $\lvert\gamma\lvert^2 = (\beta - \alpha)^2$. Here, $\alpha \text{ and }\beta \text{ are given by } e \vec{v}_{\alpha,\beta} \cdot \vec{A} /(\hbar\omega_{p})$ which are the dimensionless Volkov and Floquet interaction parameters, respectively. $\vec{v}_{\alpha,\beta}$ denotes the velocity of photoelectrons and the Fermi velocity of electrons, respectively, $e$ is the electron charge, $\vec{A}$ is the vector potential, and $\hbar\omega_{p}$ represents the pump photon energy. When the pump beam is at normal incidence, a condition almost satisfied in our experimental geometry, we consider the in-plane vector potential, $\vec{A}_{xy} = -i\vec{E}_{xy}/\omega_{p} = -iE_{xy}/\omega_{p}\times(\cos\theta_p, \sin\theta_p)$. Consequently, the sideband parameter becomes

\begin{equation}\label{gamma_expanded_2}
\gamma = \beta - \alpha = -i\frac{eE_{xy}}{\hbar\omega_{p}^{2}}\bigg[-\frac{\hbar}{m_{e}}K_{x}\cos\theta_{p} + \bigg(v_{D} - \frac{\hbar}{m_{e}}k_{D}\bigg)\cos(\theta_{p}-\theta_{k})\bigg].
\end{equation}

\begin{figure}[h]
\centering
\includegraphics[width=1\textwidth]{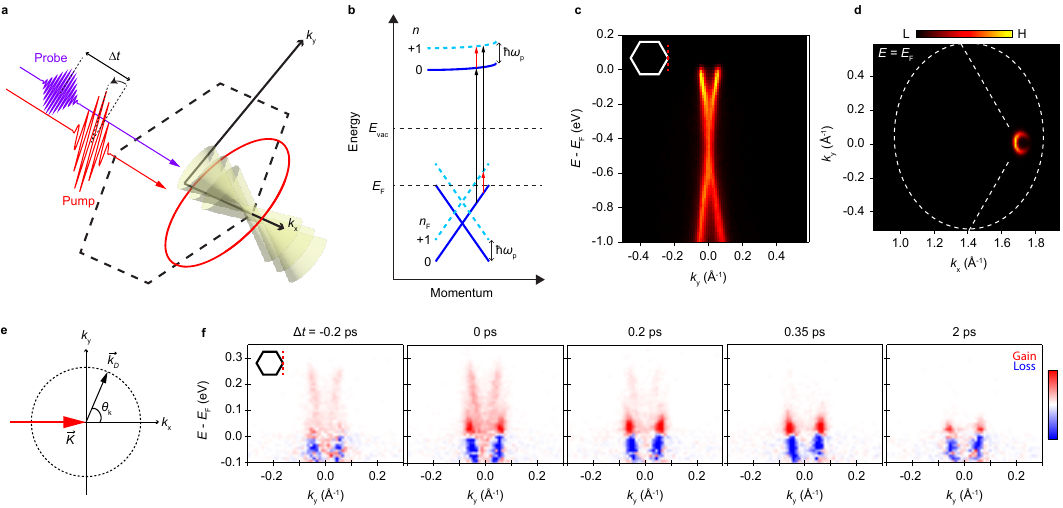}
\caption{
\textbf{Generation of a replica band in graphene via 5 um pump excitation.} \textbf{a}, Conceptual schematic of pump-probe experiments on graphene and the generation of replica bands. The red circle corresponds to the measurement window of our experiments. \textbf{b}, Illustration of the scattering between Floquet-Bloch and Volkov states \cite{Park2014,Mahmood2016,Keunecke2020}. Restricting cases to the two-photon process, $n = 1$ replica band comprises contributions from both $n_{V} = 1$ Volkov band of the photoemitted Bloch band ($n_{F} = 0$) and the photoemitted band ($n_{V} = 0$) of the $n_{F} = 1$ Floquet-Bloch band. Here, $n$, $n_{V}$, and $n_{F}$ denote the indices assigned to the replica bands resulting from Floquet-Volkov scattering, Volkov, and Floquet processes, respectively. \textbf{c}, \textbf{d}, \textit{E} – \textit{k}$_{y}$ cut (\textbf{c}) and \textit{k}$_{x}$ – \textit{k}$_{y}$ cut at \textit{E} = \textit{E}$_{F}$ (\textbf{d}) of a static spectrum of graphene. In \textbf{d}, the white dashed lines and the ellipse correspond to the Brillouin zone boundary and the measurement window, respectively. \textbf{e}, Definition of $\theta_{k}$ in \textit{k}$_{x}$ – \textit{k}$_{y}$ plane. Here, $\vec{K}$ corresponds to the vector from the origin ($\Gamma$ point) to $K$ point and $\vec{k}_{D}$ is the vector from the Dirac point (or $\vec{K}$) to the state on the Dirac cone. Note that the angle for the pump polarization, $\theta_{p}$, is defined in the same way as $\theta_{k}$ i.e. with respect to $\Gamma - K$ direction. \textbf{f}, Emergence and decay of a replica band in graphene via 5 $\mu$m pump excitation with the pump polarization angle $\theta_{p} = 12.5^{\circ}$.
}
\label{fig_1}
\end{figure}

\noindent Here, we used $\vec{v}_{\alpha,xy} = \frac{\hbar}{m_{e}} \vec{k}_{xy} = \frac{\hbar}{m_{e}} \big(K_{x}\hat{k}_{x} + \vec{k}_{D}\big)$ since the in-plane momentum of photoelectrons is conserved at the photoemission process, and the magnitude of the Fermi velocity is $\lvert\vec{v}_{\beta}\lvert = v_{D}$. Note that $\vec{k}_{D}$ corresponds to a vector from $K_{x}\hat{k}_{x}=(K_{x},0)$ to a state in the $k_{x} - k_{y}$ plane as illustrated in Fig. \ref{fig_1}e.
\par
\begin{figure}[h]
\centering
\includegraphics[width=1\textwidth]{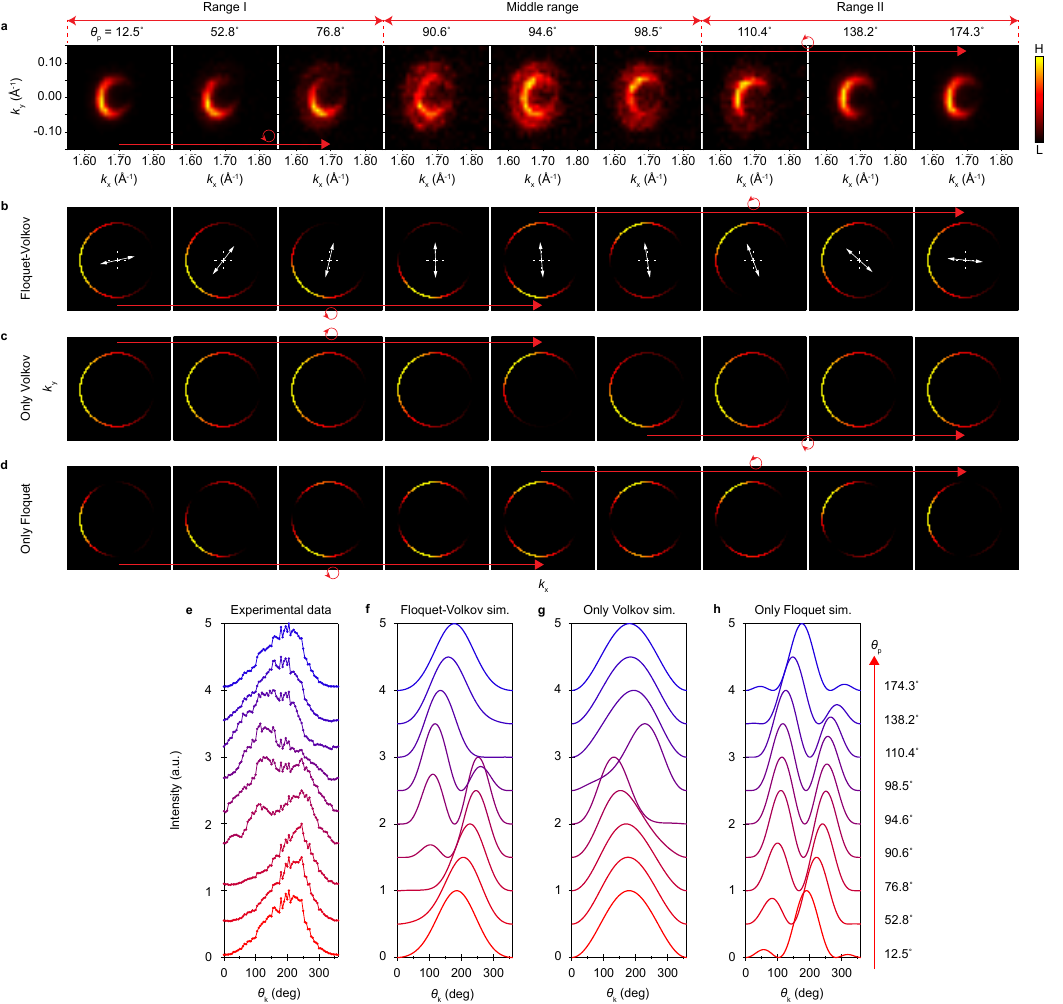}
\caption{
\textbf{Evolution of photoemission intensity  at constant energy as a function of pump polarization angle $\theta_{p}$.} \textbf{a}, ARPES spectra showing \textit{k}$_{x}$ – \textit{k}$_{y}$ cuts at $E$ – $E_{F}$ = 0.219 eV (averaged by $\pm$0.025 eV). The displayed $\theta_{p}$ range is divided into three segments: range I, middle range, and range II. \textbf{b} – \textbf{d}, Simulation results for (\textbf{b}) the scattering between Floquet-Bloch and Volkov states, (\textbf{c}) only Volkov states, and (\textbf{d}) only Floquet-Bloch states. The pump polarization directions are indicated by the white arrows in each panel. In \textbf{a} - \textbf{d}, to clearly display the replica band with a weak intensity, panels are normalized such that the color scale is set from the minimum to the maximum value within each panel. \textbf{e}, Intensity profiles extracted from \textbf{a} at each $\theta_{p}$. \textit{k}$_{x}$ – \textit{k}$_{y}$ cuts were masked to remove the outer arc, and each one was sampled at 5$^{\circ}$ intervals with an integration range of $\pm$15$^{\circ}$. Each curve was normalized. \textbf{f} – \textbf{h}, Intensity profiles for \textbf{b} – \textbf{d}, respectively.
}
\label{fig_2}
\end{figure}

 Our simulations utilizing the above model show that the experimental data can not be adequately explained by solely the Volkov states. On the other hand, considering the scattering between Floquet-Bloch and Volkov states can successfully account for the observed behaviour. The photoelectron probability of the $n=1$ replica state incorporating the non-zero incidence angle of the pump beam (see Method) selectively simulate the three distinct scenarios: Floquet-Volkov ($\beta \neq 0, \alpha \neq 0$), only Volkov ($\beta = 0, \alpha \neq 0$), and only Floquet ($\beta \neq 0, \alpha = 0$). The outcomes of these simulations are presented in Fig. \ref{fig_2}b - d. Let us begin with the only Volkov case portrayed in Fig. \ref{fig_2}c with the corresponding 1D intensity profiles shown in Fig. \ref{fig_2}g. The evolution of the spectral feature in the only Volkov case notably diverges from that observed in the experimental data of Fig. \ref{fig_2}a: firstly, the rotation directions of the arcs are opposite (experimental data: CCW, only Volkov: CW); secondly, the rotation speeds substantially differ (experimental data: gradual, only Volkov: slow to abrupt); finally, only Volkov simulation does not exhibit the simultaneous presence of two arcs at the given $\theta_{p}$ values. In the only Floquet case depicted in Fig. \ref{fig_2}d, the arc rotates CCW across the entire range, akin to the experimental data in Fig. \ref{fig_2}a. However, the emergence of the new arc transpires at a very early stage ($\theta_{p}$ = 52.8$^{\circ}$ in the range I), and three arcs are present in total at the initial and final $\theta_{p}$ values (12.5$^{\circ}$ and 174.3$^{\circ}$, respectively). These features are more apparent in the 1D intensity profiles in Fig. \ref{fig_2}h, and these behaviors diverge from the evolution of the spectral feature observed in the experimental data of Fig. \ref{fig_2}a. On the other hand, in the Floquet-Volkov case illustrated in Fig. \ref{fig_2}b, within the range I, the arc undergoes a CCW rotation as $\theta_{p}$ increases. Subsequently, in the middle range, the original arc diminishes, and a new arc at the upper left side emerges at $\theta_{p}$ = 90.6$^{\circ}$ and becomes more intense. This newly formed arc continues to rotate CCW thereafter. This progression is further supported by the intensity profiles presented in Fig. \ref{fig_2}f, indicating a close resemblance to the evolution observed in the experimental data depicted in Fig. \ref{fig_2}a. Figure \ref{fig_3} shows the peak positions of the 1D intensity profiles of the experimental data, extracted from Fig. \ref{fig_2}e, exhibiting a similar trend to that of the Floquet-Volkov case, while the only Volkov case shows a contrasting trend. Therefore, the evolution of the spectral features as the pump polarization rotates cannot be explained by the only Volkov scenario but can be effectively described by the scattering between Floquet-Bloch and Volkov states. This proves the contribution of Floquet-Bloch states to the observed replica bands in monolayer graphene.

\section{Discussion}\label{discussion}
We can derive two key insights from Eq. \ref{gamma_expanded_2} regarding (1) the strategy to minimize the Volkov parameter, $\alpha$, which differs for $K$ and $\Gamma$ electrons, and (2) the competition between Floquet-Bloch and Volkov states embedded in the trend of rotation direction of the arc. First, let us delve into the significance of the first term in Eq. \ref{gamma_expanded_2}, $-\frac{\hbar}{m_e}K_{x}\cos{\theta_p}$, which purely originates from the Volkov state. Owing to the substantial $K_{x} = 1.7$ \r{A}$^{-1}$ in graphene, this term may dominate $\gamma$ when $\cos\theta_{p}$ approaches unity. To mitigate the influence of this term, the in-plane pump polarization must be orthogonal to $\vec{K}$ (it should be polarized along the $k_{y}$ direction for our scenario). This stands in contrast to the case of the Volkov states at $\Gamma$ point, as observed in materials such as Bi$_{2}$Se$_{3}$ \cite{Mahmood2016}, where this term vanishes. In this case, for a non-zero pump incidence angle, $\vec{k_{z}} \cdot \vec{E}_{out,z}$ in $\alpha$ (resulting from the coupling between the out-of-plane components of the electric field and the photoelectron momentum $\vec{k_{z}}$) may dominate $\gamma$ \cite{Mahmood2016}. To circumvent this, s-polarization must be utilized to suppress the out-of-plane component of the electric field, $\vec{E}_{out,z}$, thereby maximizing the portion of the Floquet parameter in $\gamma$ \cite{Mahmood2016}. In the case of graphene, $\vec{k_{z}} \cdot \vec{E}_{out,z}$ is minimized because a substantial tilt of the sample is required to reach the $K$ point at (1.7 \r{A}$^{-1}$,0), causing the pump incidence angle to align nearly normal to the sample surface for our geometry. Note that, for electrons near $K$ point with a normal incidence pump beam, the first term in Eq. \ref{gamma_expanded_2} exhibits a similar effect in $\gamma$ compared to $\vec{k_{z}} \cdot \vec{E}_{out,z}$ for electrons near $\Gamma$ with a non-zero pump beam incidence angle. Both terms remain constant, independent of $\theta_{k}$ at a fixed $\theta_{p}$. In practice, even for the case of electrons near $K$, when the pump incidence angle slightly deviates from the normal incidence and the probe photon energy is substantial, as in our experiments (8.7$^{\circ}$ and 26.4 eV, respectively), $k_{z}$ can be substantial, and $\vec{k_{z}} \cdot \vec{E}_{out,z}$ may become comparable to the first term in Eq. \ref{gamma_expanded_2} (refer to the section S4 in SI). In this case, employing s-polarization which is set to be oriented perpendicular to $\vec{K}$, can minimize the impact of both terms.

\par
\begin{figure}[h]
\centering
\includegraphics[width=0.5\textwidth]{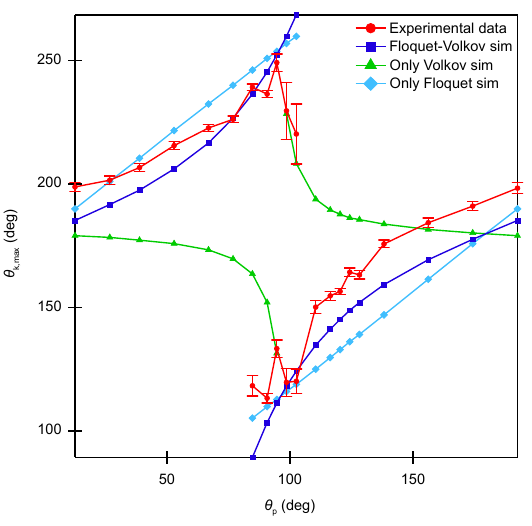}
\caption{
\textbf{The angle of the maximum intensity ($\theta_{k,max}$) as a function of pump polarization angle ($\theta_{p}$).} $\theta_{k,max}$ is plotted against $\theta_{p}$ for experimental data and three simulation cases of Floquet-Volkov, only Volkov, and only Floquet. Peak positions from the intensity profile of the experimental data (Fig. \ref{fig_2}e) were extracted by fitting a Gaussian peak with linear background. In the middle range where two arcs coexist, we extracted two peak positions by fitting two Gaussian peaks (refer to the section S1 and Fig. S6 in SI). Within this range, between $\theta_{p}$ = 98.5$^{\circ}$ and 102.5$^{\circ}$ (not shown in Fig. \ref{fig_2}a), the behaviour of the original arc in Fig. \ref{fig_2}a (or the right peak in Fig. \ref{fig_2}e) deviates from the overall trend of the experimental data shown here. The underlying cause of this deviation remains unclear and warrants further investigation.
}
\label{fig_3}
\end{figure}

 Next, the sign of the coefficient of the second term, $v_{D} - \frac{\hbar}{m_{e}}k_{D}$, signifies the competition between Floquet-Bloch and Volkov states and determines the rotation direction of the arc as $\theta_{p}$ varied. This can be understood by examining the photoelectron probability of the $n$ = 1 replica state, $P_{1} \propto \lvert M \rvert ^{2} \times \lvert \gamma \rvert ^{2}$ which is proportional to $\big[\text{sin}^{2}(\theta_{k}/2)[B\text{cos}(\theta_{p}) + A\text{cos}(\theta_{p}-\theta_{k})]^{2}\big]$, where $A = v_{D} - \frac{\hbar}{m_{e}}k_{D}$ and $B = -\frac{\hbar}{m_{e}}K_{x}$ are constants. To gain insights into how $\theta_{k,max}$ evolves as a function of $\theta_{p}$, we focus on the small $\theta_{p}$ region. Fig. \ref{fig_3} clearly show that $\theta_{k,max}$ goes up as a function of $\theta_{p}$ in this region for our data.  It can be shown that (refer to the section S5 in SI for the proof) the trend of $\theta_{k,max}$ reverses as the sign of $A$ flips: $A > 0$ leads to an increase in $\theta_{k,max}$ as $\theta_{p}$ increases, while $A < 0$ results in a decrease. Based on our experimental parameters $\big($$\frac{\hbar}{m_{e}}K_{x}$ = 2.0$\times10^{6}$ m/s, $v_{D}$ = 9.6$\times10^{5}$ m/s, and $\frac{\hbar}{m_{e}}k_{D}$ = 5.8$\times10^{4}$ m/s$\big)$, we indeed have $A > 0$ since we are in Floquet-Volkov case. The only Volkov case ($A=- \frac{\hbar}{m_{e}}k_{D} < 0$) would lead to the reverse trend which we do not observe. Hence, the different behaviour of the rotation direction of the arc between the Floquet-Volkov and the only Volkov cases stems from $v_{D} > \frac{\hbar}{m_{e}}k_{D}$, which reflects a competition between Floquet-Bloch states (which are determined by the Fermi velocity within the material $v_{D}$) and a subset of the Volkov states due to in-plane momentum with respect to the Dirac ($K$) point (which are set by the velocity $\frac{\hbar}{m_{e}}k_{D}$). 



 
\par Floquet-Bloch states are anticipated to exhibit avoided crossings along the direction perpendicular to the polarization direction of the pump beam, while remaining gapless along the polarization direction of the pump beam \cite{Syzranov2008,Lopez-Rodriguez2008,Oka2009,Lopez-Rodriguez2010,Zhou2011,Calvo2011,Wang2013,Fregoso2013,Mahmood2016}. In our experiments, the avoided crossing gaps are absent due to the pump photon energy being smaller than the gap size at the Dirac point (refer to the section S2 in SI). If the different orders were crossing, the expected gap size would be 2$\Delta$ = $\beta\omega \sim 21$ meV \cite{Syzranov2008,Lopez-Rodriguez2008,Oka2009,Lopez-Rodriguez2010,Zhou2011,Calvo2011,Wang2013,Fregoso2013,Mahmood2016}. Even if these gaps were present, it would be challenging to observe them since they would be below our energy resolution of 53 meV. However, as explained above, we can still conclusively prove the existence of Floquet-Bloch states without observing these gaps by studying the evolution of the replica bands as a function of pump polarization direction. 

\par Finally, we discuss the similarities and differences between our work and ref. \cite{Aeschlimann2021} which reported negative result for the observation of Floquet-Bloch states in graphene using trARPES. Both works utilized monolayer epitaxial graphene on SiC, albeit grown on different polytypes with distinct stacking sequences in their crystal structures (4H-SiC vs. 6H-SiC), and employed probe photon energies of 26.4 eV vs. 21.7 eV, alongside pump photon energies of 246 meV vs. 280 meV (our study vs. ref. \cite{Aeschlimann2021}). However, our measurements were performed at a lower temperature of 31 K, with a probe energy resolution of 53 meV, whereas ref. \cite{Aeschlimann2021} conducted their experiments at room temperature with 150 meV energy resolution. A potential decrease in scattering rate at a lower temperature coupled with the improved energy resolution could be responsible for the findings in our study.

\par In conclusion, we reported momentum resolved observation of Floquet-Bloch states in monolayer graphene. Measured ARPES spectra with mid-IR pump excitation shows replica bands of Dirac cone spaced by photon energy. The evolution of these replicas with the pump polarization rotation unequivocally demonstrates that they arise from the scattering between Floquet-Bloch and Volkov states. To explain these findings, we presented a simple analytical model inspired by ref. \cite{Park2014}. The approach presented here can be extended to elucidate Floquet-Bloch states at the Brillouiun zone edge in other material systems which may catalyze further advancements in Floquet engineering.


\bibliography{main_v9_0418_arXiv041924}

\newpage
\clearpage

\newpage
\clearpage

\newpage
\clearpage



\setcounter{figure}{0}
\setcounter{table}{0}
\setcounter{equation}{0}
\makeatletter 
\renewcommand{\thefigure}{M\@arabic\c@figure}
\renewcommand{\thetable}{M\@arabic\c@table}
\renewcommand{\theequation}{M\@arabic\c@equation}
\makeatother

\section{Method}\label{method}
\textbf{Materials.} A four-inch wafer monolayer epitaxial graphene was purchased from Graphenesic. The wafer was synthesized by graphitizing the Si-face 4H-SiC. To see a general description of the synthesis process, please refer to ref. \cite{Emtsev2009}. The sample utilized for measurements was cleaved from the wafer, yielding lateral dimensions of approximately 5 mm $\times$ 5 mm.  According to the factory inspection report, optical inspection confirmed a monolayer coverage of 78.8\% at the center of the wafer, accompanied by 21.2\% bilayer graphene inclusions. Note that the gapped Dirac point in Fig. \ref{fig_1}c is due to the breaking of A and B sublattice symmetry induced by interaction with the buffer layer \cite{Zhou2007}. The sample was transported in air. Prior to introduction to the measurement chamber of trARPES, it was annealed at $600^{\circ}$C for 10 minutes in a chamber at a vacuum level of low $\times 10^{-9}$ torr, which is \textit{in-situ} connected to the preparation chamber of trARPES.
\\
\par \noindent \textbf{Experimental setup.} The beam line for our trARPES setup is powered by a Yb-fiber laser (Tangerine from Amplitude). This laser operates at a repetition rate of 300 kHz and produces pulses with a center wavelength of 1030 nm (1.2 eV), duration of 135 fs, and pulse energy of 250 $\mu$J. The fundamental beam is split into a probe and a pump branch. In the probe arm, the second harmonic of the laser is generated within a harmonic box by injecting the fundamental beam into a $\beta$-BBO crystal. The resulting beam is directed into a gas-jet chamber where it is focused onto argon gas expelled from a nozzle to generate high-harmonics. As the seed beam and its harmonics propagate together, we isolate 26.4 eV beam (11th harmonic of 2.4 eV) using an XUV monochromator (McPherson Inc.), which employs gratings in an off-plane mounted configuration \cite{Sie2019}. Subsequently, the selected 26.4 eV beam is focused onto the sample via a toroidal mirror. The polarization of the probe beam is linear s-polarization and predominantly aligned along the $k_{y}$ axis (refer to Fig. S9 in SI). In the pump arm, pulses carrying 26.7 $\mu$J of energy are directed into an optical parametric amplifier (OPA) equipped with a different frequency generation (DFG) module (ORPHEUS-HP from Light Conversion) to generate the pump beam with a photon energy of 246 meV (approximately 5 $u$m in wavelength). The pump beam was directed to the sample at a nearly normal incidence angle of approximately 8.7$^{\circ}$ from the sample surface normal. The beam profile on the sample was an ellipse with the axes of $480$ $\mu$m and $440$ $\mu$m. The pulse duration (FWHM) was estimated to be about 250 fs. The fluence was around 25.4 $\mu$J/cm$^{2}$ corresponding to a peak electric field intensity of around 2.7$\times$10$^{7}$ V/m. Polarization of the pump beam was controlled using a half-waveplate with the center wavelength of 5 $\mu$m. The waveplate was purchased from VM-TIM. We employed an angle-resolved time-of-flight (ARTOF) type analyzer (ARTOF 10k from ScientaOmicron). We utilized 26-7 lens mode offering an acceptance angle of $\pm$13$^{\circ}$ and an energy window spanning $\pm$3.5\% of the center energy. The vacuum level of the measurement chamber of trARPES is mid $\times 10^{-11}$ torr. The overall setup resembles that detailed in ref. \cite{Sie2019}, but notable modifications include the different  laser with a harmonic box, high-harmonic generation (HHG) module, and OPA/DFG. In this study, we used the gas-jet method for HHG. The estimated energy resolution of the ARPES spectra obtained with the 26.4 eV probe beam is approximately 53 meV.
\\
\par \noindent \textbf{Simulations.} In the main text, for the sake of simplicity, we presented the equation for $\gamma$ by assuming a normal incidence of the pump beam (Eq. \ref{gamma_expanded_2}). However, in the simulation results depicted in Fig. \ref{fig_2}, a modified model was employed, incorporating a non-zero incidence angle of the pump beam. This modified equation for $\gamma$ is presented below:

\begin{equation}\label{gamma_non_zero_incidence_angle}
\begin{split}
\gamma & = -i\frac{e}{\hbar\omega^{2}_{p}}\bigg[-\frac{\hbar}{m_{e}}\bigg(K_{x}E_{out,xy}\text{cos}(\theta_{p,out})
       + \vec{k}_{z} \cdot \vec{E}_{out,z} \bigg) \\
       & + \bigg(v_{D}E_{t,xy}\text{cos}(\theta_{p,t}-\theta_{k}) - \frac{\hbar}{m_{e}}k_{D}E_{out,xy}\text{cos}(\theta_{p,out}-\theta_{k})\bigg)\bigg].
\end{split}
\end{equation}

\noindent Here, $E_{t,xy} = \sqrt{E_{t,x}^{2} + E_{t,y}^{2}}$ and $E_{out,xy} = \sqrt{E_{out,x}^{2} + E_{out,y}^{2}}$, which are the in-plane component of the transmitted electric field and the one outside the sample, respectively. $\vec{E}_{out,z}$ represents the out-of-plane component of the electric field intensity outside the sample. ``in-plane" and ``out-of-plane" are defined with respect to the sample surface. $\theta_{p,t}$ and $\theta_{p,out}$ denote the polarization angle with respect to the $k_{x}$ axis for the in-plane components of the transmitted electric field intensity and the one outside the sample, respectively. $\vec{k}_{z}$ denotes the $\hat{z}$ component of the momentum of photoelectrons, which can be evaluated as follows:

\begin{equation}\label{k_z}
\begin{split}
{k}_{z}  & = \sqrt{\frac{2m_{e}}{\hbar^{2}}\big(E_{probe}-E_{B}-\phi_{W}\big) - k_{xy}^{2}}
\end{split}
\end{equation}

\noindent Here, $E_{probe}$, $E_{B}$, and $\phi_{W}$ represent the probe photon energy, binding energy of the states, and work function, respectively. The term $k_{xy} = \sqrt{\big(K_{x}+k_{D}\text{cos}\theta_{k}\big)^{2} + \big(k_{D}\text{sin}\theta_{k}\big)^{2}}$ denotes the magnitude of the in-plane momentum of photoelectrons. When $\theta_{i} = 0$, Eq. \ref{gamma_non_zero_incidence_angle} reduces to Eq. \ref{gamma_expanded_2}. The parameter values used in the simulations are summarized in Table \ref{parameter_values}. For the derivation of Eq. \ref{gamma_non_zero_incidence_angle} and detailed information, please refer to the section S4 in SI.
\\
\par \noindent \textbf{Data availability.} All data needed to evaluate the conclusions in the paper are present in the paper and/or the Supplementary Information.
\\
\par \noindent \textbf{Acknowledgements.} We are grateful to M. A. Sentef, M. Eckstein, P. Werner, H. Ning, and B. Ilyas for insightful discussions. We acknowledge C. John for the detailed discussions regarding the sample. The work at MIT was supported by the US Department of Energy, BES DMSE (data acquisition, analysis, and manuscript writing) and Gordon and Betty Moore Foundation's EPiQS Initiative grant GBMF9459 (instrumentation). M.M. acknowledges the support from JST PRESTO (no. JPMJPR23HA). U.D.G, H.H., and A.R. acknowledge the support from HORIZON-MCSA-2022-DN “TIMES” (project number 101118915).
\\
\par \noindent \textbf{Author contributions.} N.G. conceived the project. D.C. and M.M. conducted trARPES experiments and simulations. D.C., M.M., U.D.G, H.H., A.R., and N.G. engaged in discussions regarding the results, and analyzed and interpreted them. D.C., M.M., D.A., B.L., and Y.S. carried out maintenance of the trARPES setup and discussions on the results. D.C., M.M., U.D.G., H.H. and N.G. wrote the manuscript. All the authors contributed to the final version of the paper. N.G. supervised the entire project. 
\\
\par \noindent \textbf{Competing interest statement.} The authors declare that they have no competing interests.

\newpage
\begin{table}[h]
\caption{Parameter values used in the simulations.}
\begin{tabular}{ |p{1cm}|p{2cm}||p{1cm}|p{2cm}| }
\toprule
$\theta_{\text{off}}$    & 12.5$^{\circ}$               & $E_{i0}$                & 2.7$\times$10$^{7}$ (V/m)   \\
\hline
$\theta_{i}$             & 8.66$^{\circ}$               & $n$                     & 5.60                        \\
\hline
$K_{x}$                  & 1.7 (\r{A}$^{-1}$)           & $E_{probe}$             & 26.4 (eV)                   \\
\hline
$k_{D}$                  & 0.05 (\r{A}$^{-1}$)          & $E_{B}$                 & -0.219 (eV)                 \\
\hline
$v_{D}$                  & 9.6$\times$10$^{5}$ (m/s)    & $\phi_{W}$              & 4.264 (eV)                  \\
\botrule
\end{tabular}
\label{parameter_values}
\end{table}
\clearpage

\end{document}